\documentclass[10pt,conference]{IEEEtran}

\usepackage{layout, calc}
\usepackage{setspace}
\usepackage{times,amssymb,amsmath,epsfig,amsthm}
\usepackage{citesort}
\usepackage[usenames]{color}

\newcommand{\myincludegraphics}[2]{\includegraphics#1#2}

\newtheorem{theorem}{Theorem}
   
   \newtheorem{corollary}{Corollary}
   \newtheorem{definition}{Definition}

\title{Two Models for Noisy Feedback in MIMO Channels}
\author{\authorblockN{Vaneet~Aggarwal}
\authorblockA{
Department of Elec. Eng.\\
Princeton University\\
Princeton, NJ 08544\\
vaggarwa@princeton.edu}
\and\authorblockN{Gajanana~Krishna}
\authorblockA{
Department of Elec. Eng.\\
Stanford University\\
Stanford, CA 94305\\
 gkrishna@stanford.edu}\and \authorblockN{Srikrishna Bhashyam}
\authorblockA{
Department of Elec. Eng.\\
Indian Institute of Technology Madras\\
Chennai, India 600036\\
skrishna@ee.iitm.ac.in} \and
\authorblockN{Ashutosh~Sabharwal}
\authorblockA{
Department of ECE\\
 Rice University\\
Houston, TX 77005\\
ashu@rice.edu} }
\date{}
\begin{document}
\maketitle
\begin{abstract}
Two distinct models of feedback, suited for FDD (Frequency Division Duplex) and TDD (Frequency Division Duplex) systems respectively, have been widely studied in the literature. In this paper, we compare these two models of feedback in terms of the diversity multiplexing tradeoff for varying amount of channel state information at the terminals. We find that, when all imperfections are accounted for, the maximum achievable diversity order in FDD systems matches the diversity order in TDD systems. TDD systems achieve better diversity order at higher multiplexing gains. In FDD systems, the maximum diversity order can be achieved with just a single bit of feedback. Additional bits of feedback (perfect or imperfect) do not affect the diversity order if the receiver does not know the channel state information.
\end{abstract}

\section{Introduction}

Channel state information to the transmitters has been extensively studied in MIMO systems \cite{narula98,Farbod_Dissertation,kim07,gamal06,steger,gaj,vaneet,jaf08,vaneetno,sk,kim08} to improve upon the diversity multiplexing tradeoff without feedback~\cite{zhength}. While earlier work often assumed noiseless feedback (possibly quantized), recent emphasis has been on studying the performance with noisy feedback~\cite{steger,gaj,jaf08,vaneetno,kim08} in single-user MIMO channels. Two distinct models of feedback have appeared. The first is that of quantized channel state information~\cite{narula98,Farbod_Dissertation,kim07,jaf08,vaneetno} which is more appropriate for asymmetric frequency-division duplex systems. The other is that of two-way training, suitable for symmetric time-division duplex systems and is the focus of study in~\cite{steger,gaj,kim08}. In this paper, we study these two systems and provide comparisons.

%In this paper, we consider the effects of varying amount of channel state information at the  transmitter and receiver for both the models of feedback. For the quantized channel state information at the transmitter, we see that although assuming perfect channel state information at the receiver and the perfect feedback link give us unbounded diversity gains as the number of feedback indices grow large, the diversity is bounded if the forward or the feedback link is imperfect. Thus, considering the noise in the system gives us bounded gain.

In this paper, we consider the effect of varying amount of channel state information at the  transmitter and the receiver for both models of feedback. It is known that assuming perfect channel state information at the receiver and a perfect feedback link gives us unbounded diversity order \cite{kim07}. Here, we find that the diversity order is bounded if feedback link is imperfect and the receiver do not know channel state information. Thus, considering the noise in the system gives us bounded diversity order.

%For the quantized feedback when the receiver knows the channel state perfectly and the feedback link is imperfect, the authors of \cite{jaf08} found that unless the errors in the feedback link decay with $\mathsf{SNR}$, the diversity cannot be improved with feedback. In \cite{vaneetno}, we found that if the forward and reverse channels are $\mathsf{SNR}$ symmetric, then we can get double the diversity at arbitrarily small multiplexing while also increasing the diversity by at most $mn$ for any general multiplexing where $m$ and $n$ are the number of transmit and receive antennas respectively. In this paper, we further extend this case to send a power controlled feedback to get a diversity gain of $mn(mn+2)$ rather than just $2mn$ before with power-controlled training.

For the quantized feedback model, where the receiver knows the channel state perfectly and the feedback link is imperfect, we find that a diversity order of $mn(mn+2)$ can be achieved with power controlled feedback rather than $2mn$ without the power-controlled feedback in \cite{vaneetno}. Further, we find that the receiver knowledge limits the diversity multiplexing performance and if the receiver needs to be trained, increasing the number of feedback levels beyond $2$ ($1$ bit) does not improve the diversity. A maximum diversity order of $mn(mn+1)$ is obtained at zero multiplexing when the receiver has to be trained with either  perfect or imperfect power-controlled feedback.

For the two-way training model, we find that unbounded diversity order is obtained if either or both of the nodes have perfect channel state information. If none of the nodes have perfect channel state information and are trained, we are limited by diversity order of $mn(mn+1)$ for zero multiplexing gain which is the same as found in the case of quantized feedback. Thus, we see that the maximum diversity order of the quantized feedback model with just $1$ bit of feedback is same as the diversity order of the two-way training model when all the imperfections in channel estimation are accounted. We further see that the matching performances above only holds for the zero multiplexing point while at general multiplexing, the two-way training model achieves higher diversity order than the quantized feedback model.

In TDD systems, we get unbounded diversity order if either of the nodes have perfect channel state information because the node with perfect channel state information can train the other well since the forward and the backward channels are the same. However, in FDD systems, this no longer holds since the forward and the backward channels are independent, and hence even if one of nodes has the channel state information of the link to it, it cannot help resolve the channel state at the other node.

The rest of the paper is outlined as follows. In Section II we give background on the two-way channel model and the diversity multiplexing tradeoff. In Section III, we summarize the diversity multiplexing tradeoff when there is no feedback link \cite{zhength}. In Section IV and V, we present the diversity multiplexing tradeoff results for the two channel models. In Section VI, we give numerical results. Section VII concludes the paper.
\section{Preliminaries}
\subsection{Two-way Channel Model}
For the single-user channel, we will assume that there are $m$ transmit antennas at the source and $n$ receive antennas at the destination. The channel input output relation is given by
\begin{equation}
Y =  HX+ W
\end{equation}
where the elements of  $H$ and $W$ are assumed to be i.i.d.\ $CN(0,1)$. The transmitter is assumed to be power-limited, such that the long-term power
is upper bounded, i.e, ${\mathbb E}\left[ X^2 \right] \leq
\mathsf{SNR}$.
Furthermore, the channel $H$ is assumed to be fixed during a fading block of $L$ consecutive channel uses, and changes from one block to another.

Since our focus will be studying feedback over noisy channels, we assume that the same multiple antennas at the transmitter and receiver are available to send feedback on an orthogonal channel.
For the feedback path, the receiver will act as a transmitter and the transmitter as a receiver. As a result, the feedback source (which is the data destination) will have $n$
transmit antennas and the feedback destination (which is the data source) will be assumed to have $m$ receive antennas. Furthermore, a block fading channel model is assumed
\begin{equation}
Y_f = H_f X_f + W_f
\end{equation}
where $H_f$ is the MIMO fading channel for the feedback link, normalized much like the forward link. The feedback transmissions are also assumed to be power-limited, that is the reverse link has a power budget of ${\mathbb E}\left[ X_f^2 \right] \leq \mathsf{SNR}$.
%Without loss of generality, we will assume a \emph{symmetry in resources}, such that $\mathsf{SNR} = \mathsf{SNR}_f$.

In this paper, we consider two types of systems: TDD and FDD. In TDD systems, the channels in the forward and the backward directions are symmetric and hence $H=H_f^\dagger$ and training can take place from the receiver to the transmitter. On the other hand, in FDD systems, the forward and the feedback path are asymmetric and hence $H$ and $H_f$ are considered independent of each other. Also, the feedback path is used to send a quantized feedback power level.

\subsection{Diversity Multiplexing Tradeoff}
In this paper, we will only consider single rate transmission where
the rate of the codebooks does not depend on the feedback index and is known to the receiver. Therefore, regardless of the feedback at the transmitter, the receiver attempts to decode the received codeword from the same codebook. Outage occurs when the transmission
power is less than the power needed for successful (outage-free) transmission \cite{zhength}.

Note that all the index mappings, codebooks, rates, powers are
dependent on $\mathsf{SNR}$. The dependence of rate on
$\mathsf{SNR}$ is explicitly given by $R = r \log \mathsf{SNR}$.\footnote{We adopt the notation
of \cite{zhength} to denote $ \buildrel.\over=$ to represent
exponential equality. We similarly use $ \buildrel.\over<$, $
\buildrel.\over>$, $ \buildrel.\over\le$, $ \buildrel.\over\ge$ to
denote exponential inequalities.}

The outage probability is the probability of outage and is formally defined in \cite{zhength,steger,vaneetit}. The system has diversity order $d(r)$ if the outage probability is $\doteq \mathsf{SNR}^{-d(r)}$ for a given multiplexing gain
$r$.  The diversity order $d(r)$  describes the achievable diversity multiplexing tradeoff.

%
%
%In point-to-point channels, outage is defined as the event that the
%mutual information of the channel, $I(X;Y | H)$ is less than the
%desired rate $R$, where $I(X;Y|H) = \log\det\left(I+\frac{P}{m}HQH^\dagger\right)$ is the mutual information of a
%point-to-point link with $m$ transmit and $n$ receive
%antennas, transmit signal to noise ratio $\mathsf{SNR}$ and input distribution Gaussian with covariance matrix $Q$~\cite{zheng03}. Since $I(X;Y|H)$ depends on transmit power, we write this dependence explicitly as $I(X;Y | H,P)$. Let $\Pi(O)$ denote the probability of outage. The
%system is said to have diversity order of $d$ if
%$\Pi(O) \doteq \mathsf{SNR}^{-d}$. The diversity multiplexing for
%the multiple users can be described as: given the multiplexing gains
%$r$ for the user, the diversity order that  can be
%achieved describes the diversity multiplexing tradeoff region.
%
%
%
%The probability of outage with rate $R$ and transmit power $P$ is
%denoted by $\Pi(R,P)\triangleq \Pi(O(R,P))$. Also let
%$U_{ H}({ R},{ P})$ be defined as the indicator function of
%$O({ R}, { P})$.
%Then, $\Pi({ R}, {P})$ is the probability of event
%$\{U_{ H}({ R},{ P})=1\}$ over the randomness of channel
%matrix ${H}$. Let $P\buildrel.\over= \mathsf{SNR}^{p}$ and ${ R} \doteq r\log
%\mathsf{SNR}$. Let $G({
%r},{ p})$ be defined as $\Pi({
%R}, { P}) \buildrel.\over= \mathsf{SNR}^{-G({
%r},{p})}$.
We will now define a function $G(r,p)$ which will be used later throughout the text. This function signifies the diversity multiplexing tradeoff for a coherent system where there is no feedback and the power constraint is $\mathsf{SNR}^p$ and the rate is $\doteq r \log(\mathsf{SNR})$.

\begin{definition}\label{outlemma}
Let $0<r< p\min(m,n)$ and $p>0$. Then, we define $G(r,p) \triangleq $
\[
 \inf \limits_{(\alpha_1,\cdots,\alpha_{\min(m,n)})\in A}
\mathop \sum
\limits_{i=1}^{\min(m,n)}(2i-1+\max(m,n)-\min(m,n))\alpha_i
\]
where $A \triangleq $
\[
\{(\alpha_1^{\min(m,n)} )| \alpha_1 \ge \ldots \alpha_{\min(m,n)}
\ge 0, \mathop \sum \limits_{i=0}^{\min(m,n)}(p-\alpha_i)^+<r\}.
\]
\end{definition}
Note that $G(r,p)$ is a piecewise linear curve connecting the points $(r,G(r,p))$=
$(kp,p(m-k)(n-k))$, $k = 0, 1, \ldots, \min(m,n)$ for fixed $m$, $n$ and $p>0$. This follows
directly from Lemma 2 of \cite{kim07}. Further, $G(r,p)=pG(\frac{r}{p},1)$.
\subsection{Summary of results}
In this paper, we will describe the diversity multiplexing tradeoff for the following cases: (1) only one-way links, (2) two-way channel with FDD-based quantized feedback reverse link, and (3) two-way channel with TDD-based reverse link. In all the three cases, we describe the effect of knowledge of channel state information at the transmitter and the receiver including imperfect knowledge due to the effect of noise in the channel while communicating the training or feedback symbols. Note that, in this paper, we ignore the time spent in training although it can be easily incorporated by seeing that $\min(m,n)$ timeslots are spent in training the receiver and the transmitter along the lines of \cite{zhength}, and replacing $r$  with $\frac{L}{L-\tau}r$ where $\tau$ is the time slots used for the training. Moreover, as pointed out in \cite{zhength}, all the antennas at the transmitter and receiver may not be needed for general multiplexing when $L$ is comparable to the number of antennas. The assumption of $L-\tau\ge m+n$ is taken in this paper so that the outage probability is of the same order as the error probability as in \cite{zhength}. Further, $r<\min(m,n)$ will be assumed throughout the paper. All the results in the paper are summarized in Table~\ref{tbl:all div-mux tradeoffs} where the FDD results are parameterized by multiplexing ($r$) and levels of feedback ($K$). For notation, $\text{R}$ represents models with perfect knowledge of channel at the receiver while $\widehat{\text{R}}$ represents those on which the receiver is trained on a noisy channel. Further, ${\text{T}}_{\text{q}}$ represents perfect quantized feedback while ${\text{T}}_c$ represents channel knowledge using a training signal from the receiver on a perfect feedback channel. $\widehat{\text{T}}_{\text{q}}$ and $\widehat{\text{T}}_c$ represents that there is an error in the feedback signal due to noise.

%\begin{figure*}
\begin{table*}
\begin{center}
\caption{Summary of diversity multiplexing Tradeoffs (ignoring training and feedback overheads).  \label{tbl:all div-mux tradeoffs}}
\begin{tabular}{|c|c|c|} \hline
Case & Main Characteristic & D-M Tradeoff \\ \hline
CSIR & Perfect Information at ${\sf R}$ & $d_{\text{CSIR}}=G(r,1)$  \\ \hline
CSI$\widehat{\text{R}}$ & No information at transmitter or receiver & $d_{\text{CSI}\widehat{\text{R}}}=G(r,1)$ \\ \hline
%CSIRT & Perfect Information at both receiver and transmitter & $d_{\text{CSIRT}}=\infty$  \\ \hline
%CSI$\widehat{\text{R}}$T & Perfect information at the transmitter & $d_{\text{CSI}\widehat{\text{R}}\text{T}}=\infty$\\
%\hline
CSIR$\text{T}_{\text{q}}$ & Perfect quantized CSI at transmitter & $d_{\text{CSI}\text{R}\text{T}_\text{q}}(r,K) = G(r,1+d_{\text{CSIR}\text{T}_\text{q}}(r,K-1))$,\\& Perfect CSI at receiver & $d_{\text{CSI}\text{R}\text{T}_\text{q}}(r,0)=0$ \\ \hline
CSI$\widehat{\text{R}}\text{T}_{\text{q}}$& Perfect quantized CSI at transmitter & $d_{\text{CSI}\widehat{\text{R}}\text{T}_\text{q}}(r,K)=G(r,1+G(r,1))$\\ \hline
CSIR$\widehat{\text{T}}_{\text{q}}$ & Noisy quantized CSI at transmitter & ${d}_{\text{CSI}\text{R}\widehat{\text{T}}_\text{q}}(r,K) = \min\left(d_{\text{CSI}\text{R}\text{T}_\text{q}}(r,K),\max_{q_j\le1+d_{\text{CSI}\text{R}\text{T}_\text{q}}(r,j)}
\right.$\\& Perfect CSI at receiver &$\left.\min_{i=1}^{K-1}(mn((q_i)^+-(q_{i-1})^+)+d_{\text{CSI}\text{R}\text{T}_\text{q}}(r,i)) \right)$  \\ \hline
CSI$\widehat{\text{R}}\widehat{\text{T}}_{\text{q}}$ & No Genie-aided information with quantized feedback & $d_{\text{CSI}\widehat{\text{R}}\widehat{\text{T}}_q}(r,K)=G(r,1+G(r,1))$\\ \hline
CSIR$\text{T}_{\text{c}}$ & Noise-free training channel to transmitter & $d_{\text{CSIR}\text{T}_c}=\infty$\\& Perfect CSI at receiver & \\ \hline
CSI$\widehat{\text{R}}\text{T}_{\text{c}}$& Noiseless channel to transmitter & $d_{\text{CSI}\widehat{\text{R}}\text{T}_c}=\infty$\\ \hline
CSIR$\widehat{\text{T}}_{\text{c}}$ & Noisy channel to transmitter & ${d}_{\text{CSIR}\widehat{\text{T}}_\text{c}} = \infty$\\& Perfect CSI at receiver &  \\ \hline
CSI$\widehat{\text{R}}\widehat{\text{T}}_{\text{c}}$ & No Genie-aided information with &$d_{\text{CSI}\widehat{\text{R}}\widehat{\text{T}}_c}=mn+G(r,mn)$\\& symmetric two-way channel & \\ \hline
\end{tabular}
\end{center}
\end{table*}
%\end{figure*}

\section{No feedback channel}

%(VANEET, CSIRT AND CSI$\hat{R}$T ARE SAME AS SECTION V.B AND V.C. ALSO CSIRT AND CSI$\hat{R}$T ARE NOT REALLY NO FEEDBACK CASES. SO REMOVE SECTIONS II.C AND II.D).

In this section, we will consider that there is no feedback link from the receiver to the transmitter and all the transmissions are one-way from the transmitter to the receiver. We consider the cases of perfect channel state information and no channel state information at the receiver.

\subsection{CSIR: Perfect Channel State Information at the receiver}
Suppose that the receiver has a perfect channel estimate $H$ while the transmitter does not know $H$. The transmitter sends signal at rate $R$ assuming that it will be possible for the decoder to decode. The fading blocks in which the receiver is not able to decode results in outage. The diversity multiplexing tradeoff in this case is given as follows.
\begin{theorem}[\cite{zhength}] The diversity multiplexing tradeoff curve for the case of perfect CSIR is given by $d_{\text{CSIR}}=G(r,1)$ for $r<\min(m,n)$.
\end{theorem}
\subsection{CSI$\widehat{\text{R}}$: Estimated CSIR}
Suppose that both the receiver and the transmitter have no channel state information. In this case, the transmitter first trains the receiver at a power level of $\mathsf{SNR}$ and then sends data at rate $R$. Using this training based scheme, the diversity multiplexing tradeoff is given as follows.
\begin{theorem}[\cite{zhength}] The diversity multiplexing tradeoff curve for the case of estimated CSIR is given by $d_{\text{CSI}\widehat{\text{R}}}=G(r,1)$ for $r<\min(m,n)$.
\end{theorem}
%\subsection{CSIRT: Perfect Channel State Information at both transmitter and receiver}
%Suppose that both the receiver and the transmitter have perfect knowledge of the channel. The transmitter can use rate adaptation and always send at a rate so that the receiver will be able to decode. In this case, there is no outage and thus the diversity order is infinite.
%
%Note that infinite diversity order can also be achieved in this scenario by using power control with codebooks of a fixed rate. The transmitter sends data at rate $R$ using a power control which assigns power of the order of $\mathsf{SNR}$ divided by the probability of getting channel $H$ when the channel is $H$ as in \cite{kim08}.
%
%\begin{theorem}Suppose that $r<\min(m,n)$. Then, the diversity order for the case of perfect CSIR with perfect CSIT is $d_{\text{CSIRT}}=\infty$.
%\end{theorem}
%\subsection{CSI$\widehat{\text{R}}$T: Estimated CSIR with perfect CSIT}
%Suppose that the transmitter has perfect knowledge of the channel. In this case, the transmitter first trains the receiver using the power level as described in the case of CSIRT and then sends the data at rate $R$ at the same power level. This scheme similarly leads to infinite diversity order.
%\begin{theorem}
%Suppose that $r<\min(m,n)$. Then, the diversity multiplexing tradeoff for the case of perfect CSIT is $d_{\text{CSI}\widehat{\text{R}}\text{T}}=\infty$.
%\end{theorem}
We note in this section that the channel state information at the receiver does not affect the diversity order calculations, i.e., the receiver with no knowledge of the channel state information can be trained to give the same diversity multiplexing tradeoff as the receiver with perfect knowledge of channel state information at the receiver.

%We note in this section that it is the channel state information at the transmitter that matters in diversity order calculations. The receiver with no knowledge of the channel state information can be trained to give the same diversity multiplexing tradeoff as the receiver with perfect knowledge of channel state information at the receiver.
\section{FDD systems: Quantized Feedback channel}
In this section, we will consider that there is a feedback channel on which quantized feedback coded as power levels can be sent. There are $K$ power levels on the feedback link. $K=1$ represents no feedback, and hence we will consider $K>1$ in this section. We will consider two cases: (1) when the receiver has perfect channel state information and (2) when it has no channel state information and is trained by the transmitter. For both of these cases, we also consider two cases representing whether the feedback signal transmitted from the receiver is received perfectly or is corrupted by feedback channel noise. A non-power controlled feedback scheme was used in \cite{vaneetno} to get an increase in diversity order with imperfect feedback while, in this paper, we extend the results to a power controlled feedback scheme.
\subsection{CSIR$\text{T}_q$: Perfect CSIR with quantized CSIT}\label{csirtq}
Suppose that the receiver has perfect channel state information, and the feedback link to the transmitter is also perfect and thus the feedback signal transmitted from the receiver is received perfectly at the transmitter. In this case, the receiver first decides a feedback index based on the channel and hence decides the power level. The receiver then sends this power level to the transmitter which is received perfectly and hence the transmitter decides on a power level based on what it decoded. Using this scheme, the following diversity multiplexing tradeoff can be obtained.
\begin{theorem}[\cite{kim07}]Suppose that $K\ge1$ and $r<\min(m,n)$. Then, the diversity multiplexing tradeoff for the case of perfect CSIR with noiseless quantized feedback is given as $d_{\text{CSIR}\text{T}_q}=B^*_{m,n,K}(r)$, where $B^*_{m,n,K}(r)$ is defined recursively as $B^*_{m,n,K}(r) =
G(r,1+B^*_{m,n,K-1}(r))$, $B^*_{m,n,0}(r)=0$.
\end{theorem}
\subsection{CSI$\widehat{\text{R}}\text{T}_q$: Estimated CSIR with perfect CSI$\text{T}_q$}
In this subsection, we consider that the receiver does not know channel state information while the feedback link from the receiver to the transmitter is perfect. The transmission scheme in this scenario follows three rounds. In the first round, the transmitter sends a training signal to the receiver at power level of $\mathsf{SNR}$. In the second round, the receiver uses the estimated channel state information to choose a feedback index for the transmitter which is received perfectly at the transmitter. In the third round, the transmitter chooses a power level based on what it decoded to first train the receiver and then transmits data at the same power level. Using this mechanism, the following diversity multiplexing tradeoff can be obtained.
\begin{theorem}[\cite{vaneetit}]
Suppose that $K>1$ and $r<\min(m,n)$. Then, the diversity multiplexing tradeoff is given by $d_{\text{CSI}\widehat{\text{R}}\text{T}_q}=G(r,1+G(r,1))$.
\end{theorem}
\begin{corollary} The maximum diversity order that can be achieved in this case for $r\to 0$ is $mn(1+mn)$ which is $(mn)^2$ greater than the diversity order without feedback.
\end{corollary}
\subsection{CSIR$\widehat{\text{T}}_q$: Perfect CSIR with noisy CSI$\text{T}_q$}
In this subsection, the receiver has perfect channel state information while the feedback link is not perfect. The signal transmitted from the receiver is not received perfectly at the transmitter. We consider that the feedback is also power controlled and MAP estimation is done to decode the power levels. In this scenario, the receiver calculates and transmits the power level to the transmitter which then sends data at a power level based on what it decoded. Using this mechanism, the following diversity multiplexing tradeoff can be obtained.
\begin{theorem}[\cite{vaneetit}]
Suppose that $K>1$ and $r<\min(m,n)$. Then, the diversity multiplexing tradeoff  is given by
 $d_{\text{CSIR}\widehat{\text{T}}_q}\doteq\min(B_{m,n,K}(r),
\max_{q_j\le1+B_{m,n,j}(r)}\min_{i=1}^{K-1}(mn((q_i)^+-(q_{i-1})^+)+B_{m,n,i}(r)))$.
\end{theorem}
\begin{corollary}
The diversity multiplexing tradeoff with one bit of imperfect feedback is same as the diversity multiplexing tradeoff with one bit of perfect feedback. In other words, for $K=2$, $d_{\text{CSIR}\widehat{\text{T}}_q}=G(r,1+G(r,1))$.
\end{corollary}
\begin{corollary}
For $K\to\infty$ and $r\to 0$, the maximum diversity order that can be obtained in this case is $mn(mn+2)$ which is $(mn)^2$ more as compared to feedback scheme in \cite{vaneetno}.
\end{corollary}
\subsection{CSI$\widehat{\text{R}}\widehat{\text{T}}_q$: Estimated CSIR with noisy CSI$\text{T}_q$}
In this subsection, the receiver does not know the channel state information. Moreover, the feedback link from the receiver to the transmitter is imperfect. The transmission scheme in this scenario follows three rounds. In the first round, the transmitter sends a training signal to the receiver at power level of $\mathsf{SNR}$. In the second round, the receiver uses the estimated channel state information to obtain a feedback index for the transmitter which is received imperfectly at the transmitter. In the third round, the transmitter uses the power level based on what it decoded to first train the receiver and then transmits data at the same power level. Using this mechanism, the following diversity multiplexing tradeoff can be obtained.
\begin{theorem}[\cite{vaneetit}]
Suppose that $K>1$ and $r<\min(m,n)$. Then, the diversity multiplexing tradeoff is given by $d_{\text{CSI}\widehat{\text{R}}\widehat{\text{T}}_q}=G(r,1+G(r,1))$.
\end{theorem}
\begin{corollary}
The diversity multiplexing tradeoff with imperfect feedback and receiver training is same as the diversity multiplexing tradeoff with one bit of perfect feedback with perfect knowledge of channel state information at the receiver.
\end{corollary}

In this section, we note that the knowledge of channel state information at the receiver also matters for diversity order. The diversity order obtained after training the receiver is less than that when the channel state information is perfectly known to the receiver. We further note that if the receiver does not know the channel state information, the diversity order is the same, irrespective of the feedback link being perfect or noisy. Since the forward and the backward channels were independent, if one of the nodes know the channel of the link to it, it cannot train the other node to its corresponding channel well (since it does not know of the channel to the other node) thus resulting in bounded gains.

\section{TDD systems: Symmetric channel}
In this section, we will consider that there is a symmetric feedback channel on which training signals can be sent. We will consider two cases: (1) when the receiver has perfect channel state information and (2) when it has no channel state information and is trained by the transmitter. For both of these cases, we also consider two cases representing whether the training signal transmitted from the receiver is received perfectly or is corrupted by additive white noise in the feedback channel.

\subsection{CSIR$\text{T}_c$: Perfect CSIR with CSIT obtained by perfect training}
Suppose that the receiver knows the channel state information perfectly and the transmitter can receive training symbols perfectly. The receiver can train the transmitter perfectly and thus, we get infinite diversity order.
\begin{theorem}Suppose that $r<\min(m,n)$. Then, the diversity order for the case of perfect CSIR with noiseless continuous feedback signal is $d_{\text{CSIR}\text{T}_c}=\infty$.
\end{theorem}

\subsection{CSI$\widehat{\text{R}}\text{T}_c$: Estimated CSIR with perfect CSI$\text{T}_c$}
Suppose that the feedback channel is perfect and neither the transmitter nor the receiver know any channel state information. In this scenario, a training symbol is sent from the receiver letting the transmitter know the exact channel. Let the channel realization be $H$, and the non-zero eigenvalues of $HH^\dagger$ be $\lambda_1, \cdots, \lambda_{\min(m,n)}$. Let $\lambda_i\doteq\mathsf{SNR}^{-\alpha_i}$. Then, the transmitter trains the receiver using power level $\mathsf{SNR}$ divided by the probability of ${\mathbf\alpha}=(\alpha_1,\cdots,\alpha_{\min(m,n)})$ as in \cite{kim08}. This is followed by sending the data at the same power level. This scheme achieves infinite diversity order in this case. Hence,

%After that, the scheme for CSI$\widehat{\text{R}}$T can be used to get infinite diversity order.
\begin{theorem}
Suppose that $r<\min(m,n)$. Then, the diversity multiplexing tradeoff for the case of perfect CSI$\text{T}_c$ with receiver training is $d_{\text{CSI}\widehat{\text{R}}\text{T}_c}=\infty$.
\end{theorem}
%\vspace{-.2in}

\subsection{CSIR$\widehat{\text{T}}_c$: Perfect CSIR with estimated CSI$\text{T}_c$}
In this subsection, we consider that the receiver has perfect channel state information while the symmetric feedback link is imperfect and the received feedback signal is not the
 same as transmitted due to the action of noise. In this scenario, the receiver sends a power controlled symbol to the transmitter for training. Consider that the receiver quantizes
  the power levels into $\{0,1,\cdots,K-1\}$. This power division is same as is done in \cite{kim07}. When the channel is $H$, the receiver finds the eigenvalues of $HH^\dagger$ as
  $\lambda_i=$ $i^{th}$ eigenvalue of $HH^\dagger$. Let $\lambda_i\doteq\mathsf{SNR}^{-\alpha_i}$ and let $\alpha_m=\min(\alpha_i)$. The receiver encodes the training 
  symbol at power level $\mathsf{SNR}^{\alpha_m+\frac{i+1}{2K}}$ to communicate power level $i$. The transmitter estimates the power level by observing the received power and sends data at the power level as in Section \ref{csirtq}.
   It can be seen that this scheme results in diversity multiplexing tradeoff equivalent to $K$ levels of perfect feedback and hence as $K$ grows large enough, we get infinite diversity.
%\vspace{-.1in}
%The receiver estimates the channel and sends a power controlled data based on its estimated channel. The diversity obtained by this scheme is infinite.
\begin{theorem}
Suppose that $r<\min(m,n)$. Then, the diversity order for the case of perfect CSIR with training the transmitter is $d_{\text{CSIR}\widehat{\text{T}}_c}=\infty$.
\end{theorem}
\vspace{-.1in}
\subsection{CSI$\widehat{\text{R}}\widehat{\text{T}}_c$: Estimated CSIR with estimated CSI$\text{T}_c$}
In this subsection, we consider that neither the receiver nor the transmitter know the exact channel. The feedback channel and the forward channel are symmetric and both are imperfect. In this scenario, the communication proceeds by the following two rounds. In the first round, the receiver trains the transmitter with a power level of $\mathsf{SNR}$. In the second round, the transmitter uses the estimated channel to select a power level on which the receiver is trained and the data is transmitted.
\begin{theorem}
Suppose that $r<\min(m,n)$. Then, the diversity multiplexing tradeoff is given by $d_{\text{CSI}\widehat{\text{R}}\widehat{\text{T}}_c}=mn+G(r,mn)$.
\end{theorem}
\begin{corollary} The maximum diversity order as $r\to 0$ is $mn(1+mn)$ which is same as the corresponding diversity order in the case of CSI$\widehat{\text{R}}\widehat{\text{T}}_q$. Thus, when all the noise in both the direct and the feedback channel are accounted for, the maximum diversity order is same in both the symmetric and asymmetric channel models.
\end{corollary}
In this section, we note that if the transmitter or the receiver knows perfect channel state information, unbounded diversity gains are obtained. Since the forward and the backward channels were assumed the same, the node knowing the channel state perfectly can train the other using power controlled training and result in unbounded diversity order. However, if none of the nodes know the channel state information perfectly, we get bounded diversity order.
\vspace{-.1in}
\section{Numerical Results}
We now see the diversity multiplexing tradeoff for the different scenarios. In Figure 1, $m=n=2$. For the quantized feedback model, one bit of feedback is assumed. We can see the seven diversity multiplexing tradeoffs in the figure, and the remaining three cases give infinite diversity. When the feedback link is perfect, the FDD system gives bounded diversity order while the TDD system gives unbounded diversity order. When all the imperfections are accounted for, the symmetric channel in TDD gives better diversity multiplexing performance than the asymmetric channel in FDD systems although the two meet as $r\to 0$.
\vspace{-.15in}
\begin{figure}[htbp]
\centering \myincludegraphics[height=5cm]
{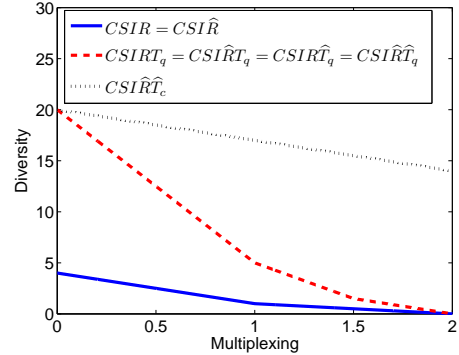}\caption{The diversity multiplexing tradeoff for $m=n=2$ and 1 bit of feedback.}\label{fig:noisyfeed}
\end{figure}
\vspace{-.2in}

\section{Conclusions}
%Channel state information at the receiver and the transmitter is imperfect due to noise. Inspired by this fact, we compare the two channel models in the presence of
%errors in terms of the diversity multiplexing tradeoff. We find in TDD systems that if one of the links is perfect, the diversity is unbounded which is not true in FDD systems. However when all the imperfections are accounted, FDD and TDD systems give same diversity at low multiplexing gains.
In this paper, we compare the FDD and TDD channel feedback models in the presence of errors. We find in TDD systems that if one of the nodes have perfect channel state information, the diversity order is unbounded which is not true in FDD systems. However when all the imperfections are accounted, FDD and TDD systems give same diversity order at arbitrarily low multiplexing gains while the TDD model achieves higher diversity order than FDD model for general multiplexing gains.

\end{document}